\documentclass[sigconf]{acmart}

\usepackage{tabularx}
\usepackage{xspace}
\usepackage{listings}
\usepackage[underline=false]{pgf-umlsd}
\usepackage{msc}
\usepackage[htt]{hyphenat}
\usepackage{hyperref}
\hypersetup{breaklinks=true,hidelinks=true}
\usepackage{booktabs}
\usepackage{multirow}
\usepackage{color, colortbl}
\usepackage{longtable}
\usepackage[ruled,linesnumbered]{algorithm2e}
\usepackage{listings}
\usepackage{svg}
\usepackage[binary-units]{siunitx}
\usepackage{makecell}

\RestyleAlgo{ruled}

\def \showcomments {Show comments}
\ifdefined\showcomments
\usepackage{todonotes}
\usepackage{letltxmacro}
\LetLtxMacro{\todonote}{\todo}
\renewcommand{\todo}[2][]
{\todonote[inline, caption={#2}, size=\footnotesize, #1]
{\renewcommand{\baselinestretch}{0.5}\selectfont#2\par}}
\else
\usepackage[disable]{todonotes}
\fi
\newcommand{\taggedpara}[1]{\noindent\textbf{#1.}}

\newcommand{\sys}{\texttt{RA-WEBs}\xspace}

\definecolor{mygreen}{rgb}{0,0.6,0}

\lstdefinestyle{gostyle}{ 
  backgroundcolor=\color{white},   %
  basicstyle=\ttfamily\scriptsize,        %
  breakatwhitespace=false,         %
  breaklines=true,                 %
  captionpos=b,                    %
  commentstyle=\color{mygreen}\ttfamily,    %
  deletekeywords={...},            %
  escapeinside={\%*}{*)},          %
  extendedchars=true,              %
  frame=single,	                   %
  keepspaces=true,                 %
  keywordstyle=\color{blue}\ttfamily,       %
  language=Go,                 %
  morekeywords={*,...},            %
  numbers=left,                    %
  numbersep=5pt,                   %
  numberstyle=\tiny\color{gray}\ttfamily,   %
  rulecolor=\color{black},         %
  stringstyle=\color{red}\ttfamily,
  showspaces=false,                %
  showstringspaces=false,          %
  showtabs=false,                  %
  tabsize=2,	                   %
  title=\lstname                   %
}

\title{RA-WEBs: Remote Attestation for WEB services }

\author{Kosei Akama}
\email{akama@keio.jp}
\orcid{0009-0001-0762-4014}
\affiliation{%
  \institution{Keio University}
  \city{Fujisawa}
  \state{Kanagawa}
  \country{Japan}
}

\author{Yoshimichi Nakatsuka}
\email{yoshimichi.nakatsuka@inf.ethz.ch}
\orcid{0009-0003-3139-097X}
\affiliation{%
  \institution{ETH Zurich}
  \city{Zurich}
  \country{Switzerland}
}

\author{Korry Luke}
\email{koluke@sfc.wide.ad.jp}
\orcid{0000-0003-3934-0611}
\affiliation{%
  \institution{Keio University}
  \city{Fujisawa}
  \state{Kanagawa}
  \country{Japan}
}

\author{Masaaki Sato}
\email{masaaki.sato@tokai.ac.jp}
\orcid{0009-0007-3280-1207}
\affiliation{%
  \institution{Tokai University}
  \city{Shinagawa}
  \state{Tokyo}
  \country{Japan}
}

\author{Keisuke Uehara}
\email{kei@sfc.keio.ac.jp}
\orcid{0000-0002-3385-6150}
\affiliation{%
  \institution{Keio University}
  \city{Fujisawa}
  \state{Kanagawa}
  \country{Japan}
}

\begin{document}

\newcommand{\verifier}{\mathtt{verifier}}
\newcommand{\ta}{\mathtt{ta}}
\newcommand{\ctmonitor}{\mathtt{ctmonitor}}
\newcommand{\ctlog}{\mathtt{ctlog}}

\newcommand{\user}{\mathtt{user}}
\newcommand{\service}{\mathtt{service}}
\newcommand{\mlist}{\mathtt{LIST}}
\newcommand{\domain}{\mathtt{domain}}
\newcommand{\pk}{\mathtt{pk}}
\newcommand{\sk}{\mathtt{sk}}
\newcommand{\evidence}{\mathtt{evidence}}
\newcommand{\tainfo}{\mathtt{INFO}_\ta}
\newcommand{\loggerinfo}{\mathtt{INFO}_\logger}
\newcommand{\cert}{\mathtt{cert}}
\newcommand{\object}{\mathit{object}}
\newcommand{\entity}{\mathit{entity}}
\newcommand{\code}{\mathtt{code}}
\newcommand{\activate}{\mathtt{activate}}
\newcommand{\valid}{\mathtt{valid}}
\newcommand{\precert}{\mathtt{precert}}
\newcommand{\sct}{\mathtt{sct}}
\newcommand{\violation}{\mathtt{violation}}
\newcommand{\violationlogs}{\mlist_{\violation}}
\newcommand{\rv}{\mathtt{rv}}

\begin{abstract}
Data theft and leakage, caused by external adversaries and insiders, demonstrate the need for protecting user data.
    Trusted Execution Environments (TEEs) offer a promising solution by creating secure environments that protect data and code from such threats.
    The rise of confidential computing on cloud platforms facilitates the deployment of TEE-enabled server applications, which are expected to be widely adopted in web services such as privacy-preserving LLM inference and secure data logging.
    One key feature is Remote Attestation (RA), which enables integrity verification of a TEE.

However, \emph{compatibility} issues with RA verification arise as no browsers natively support this feature, making prior solutions cumbersome and risky.
    To address these challenges, we propose \sys (\textbf{R}emote \textbf{A}ttestation for \textbf{Web} \textbf{s}ervices), a novel RA protocol designed for high compatibility with the current web ecosystem.
    \sys leverages established web mechanisms for immediate deployability, enabling RA verification on existing browsers.
    We conduct a comprehensive security analysis, demonstrating \sys's resilience against various threats.
    Our contributions include the \sys proposal, a proof-of-concept implementation, an in-depth security analysis, and publicly available code for reproducible research.
\end{abstract}

\maketitle

\section{Introduction}
\label{sec:intro}

In recent years, data theft and leakage have emerged as a significant concern.
    According to the Identity Theft Resource Center's annual data breach report, a total of 3,122 data breaches were reported in the US in 2023~\cite{ITRC-databreach}.
    Furthermore, the global average cost of a data breach was reported to be approximately 4.88 million USD in 2024~\cite{IBM-databreach}.
    The majority of the cases occur when businesses are targeted by external adversaries.
    However, there are also cases where an attack could occur from \emph{within} the service operator.
    For instance, services themselves could be malicious and leak data to external parties.
    Additionally, disgruntled employees may secretly steal and sell data, motivated by financial gains, retaliation, or other personal reasons.
    Notably, the cost of a data breach caused by an insider threat is substantial, amounting to 4.99 million USD~\cite{IBM-databreach}.

This highlights the need for protecting user data from even the service provider itself.
    One promising solution to this problem is to run such services within Trusted Execution Environments (TEEs)~\cite{anati2013innovative,hoekstra2013using,TrustZone,SEV,SEV-SNP,TDX,lee2020keystone}.
    TEE is a security primitive that creates secure environments, protecting data and code from various threats, including malicious service operators. 
    A key feature of a TEE is Remote Attestation (RA), which allows TEEs to prove to a remote party the validity of the TEE and the program running within it in a cryptographically secure manner.
    Cloud service platforms have taken notice of the features offered by TEEs, which led to the introduction of confidential computing~\cite{Alibaba-CC,Tencent-CC,AWS-CC,GCP-CC,Azure-CC,Oracle-CC,IBM-CC}, enabling services to easily access TEEs.

The unique properties of TEEs and the widespread availability of confidential computing on cloud services have led to a multitude of proposals for server applications that leverage TEEs.
    Given such trends, TEE-enabled server-side applications are anticipated to gain widespread adoption in the field of web services.
    Such services include, but are not limited to, the following: 
\begin{itemize}
        \item {\bf Privacy-preserving LLM inference:} Protecting users' LLM prompts~\cite{Continuum}.
        \item {\bf Privacy-Friendly DNN training:} Protecting collected training data for deep learning systems.
        \item {\bf Privacy-preserving Questionnaire System:} Questionnaire systems are designed to limit the number of responses from individual users while protecting user privacy~\cite{questionairesystem}.
        \item {\bf Non-Repudiable Logger:} Storing user-web server communications for auditing and accountability~\cite{aublin2018libseal}.
\end{itemize}

\textbf{The problem: }
However, \emph{compatiblity} becomes a major obstacle for users of TEE-enabled web services, especially regarding RA.
    This is because current web mechanisms (especially browsers) lack support for RA verification.

Many prior works have realized this issue and have proposed a multitude of approaches, one popular approach being to require users to install additional software, such as specialized browser extensions.
    One downside of this approach is user friction.
        It is known that users tend to hesitate to install software packages; a survey revealed that a significant proportion of users are deterred by additional software installation, with 53.1\% of respondents stating that they would either stop or consider stopping using a service if such installation is required~\cite{UserHateInstalling:online}.
    Moreover, installing the additional packages may cause information leakage~\cite{arcanum2024xie}, reinforcing users' reluctance to install additional software.

Some may argue that we should wait until RA verification is integrated into browsers, allowing users to leverage RA without the inconvenience of software installation.
    However, there is no guarantee that all browsers will support RA because this strongly depends on the browser vendor's decision.
Standardization via the World Wide Web Consortium (W3C) may help lower the barrier when introducing RA to a browser.
    Regardless, standardization faces several challenges.
    One key challenge is that standardization efforts do not always result in a finalized specification due to a multitude of factors, including conflicting interests and disagreements between the involved parties.
    Moreover, standardization bodies cannot force browser vendors to follow specifications, as they may simply choose to ignore them.
        Similar decisions are commonly seen in practice; for instance, Speech Recognition and Remote Playback APIs are supported in Chrome and Safari but not in Firefox~\cite{SpeechRecognition58:online, RemotePlayback:online}.

Given these challenges, it is unlikely that waiting for the browser to integrate RA verification will provide a timely solution.
    Therefore, building an RA verification system that can be \emph{deployed immediately} is necessary, especially given that a wide range of web services utilizing TEEs have already begun to appear.
Motivated by the aforementioned challenges surrounding RA in the web context, we pose the following question:
\emph{Can we create an immediately deployable RA protocol for web services that is compatible with existing browsers?}

\textbf{Our approach: }
In this work, we answer the question in the affirmative, by proposing \sys (\textbf{R}emote \textbf{A}ttestation for \textbf{Web} \textbf{s}ervices), a novel RA protocol for TEE-enabled web services, which enables users to verify RA proofs without the need of installing any additional software.

We design \sys so that it is \emph{highly compatible} with the current web ecosystem.
    The proposed system realizes this by utilizing a carefully selected set of known and well-established web mechanisms, such as CA, Web PKI, and Certificate Transparency.
    However, combining these mechanisms is not as straightforward as one might assume.
    Another highlight of this work is identifying the challenges during the design and providing meaningful solutions to them. %

Another important contribution of this work is the implementation of \sys, allowing it to be \emph{deployed immediately}, thus enabling users to conduct RA verifications with their existing browsers.
    This is thanks to the use of well-established web mechanisms in \sys.
    We also conduct an extensive performance evaluation of our unoptimized proof-of-concept (PoC) implementation to assess its performance.

Ensuring the \emph{security} of \sys is also essential to this work.
    To this end, we conducted an extensive security analysis of \sys.
    We listed out various attacks that are possible on \sys and show that \sys is secure against such threats.

To contribute to \emph{reproducible research}, source code for all \sys components
    is available at \cite{AnonymousGithubOnline}.
    We hope this facilitates further research into this area and web systems that utilize TEEs to take advantage of \sys.

In summary, the anticipated contributions of this study are: 
\begin{enumerate}
    \item The proposal of \sys, which is designed to be highly compatible with the existing web ecosystem. 
    \item A PoC implementation of \sys, showcasing its immediate deployability in the real world.
    \item An extensive evaluation of \sys, including in-depth security and compatibility analysis.
    \item Publicly available \sys source code 
        to facilitate reproducible research.
\end{enumerate}    

\section{System and Threat Model}
\label{sec:system}

\begin{figure}[t]
  \centering
  \includegraphics[width=80mm]{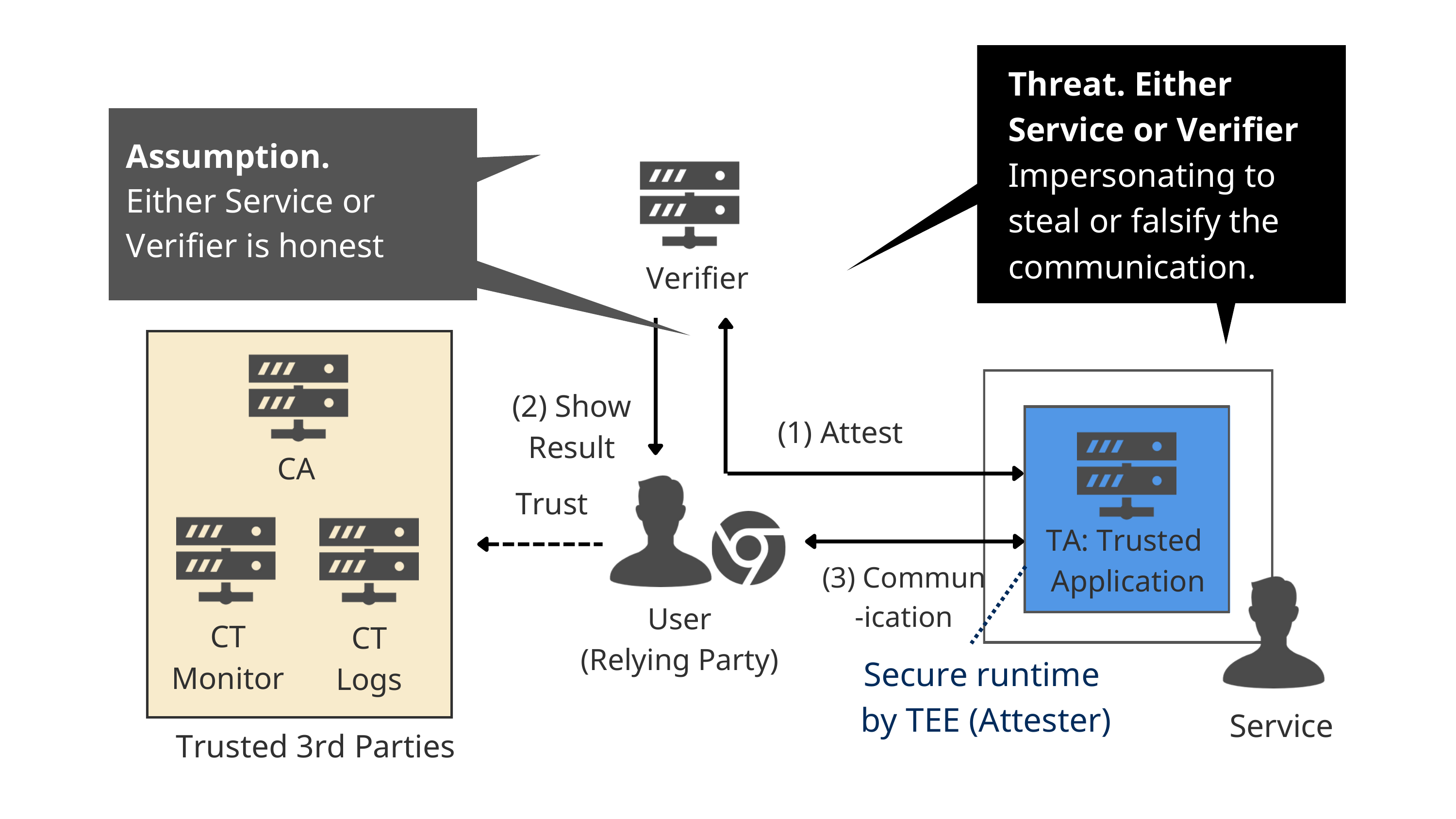}

  \vspace{-4mm}
  \caption{System \& Threat Model of {\sys} }
  \label{fig:system}
  \vspace{-4mm}

\end{figure}

Figure~\ref{fig:system} shows the overall system.
    {\sys} consists mainly of the following six entities:

\begin{itemize}
    \item \textbf{User:} Utilizes the service. %
    \item \textbf{Service:} Operates or manages TA (e.g., web service).
    \item \textbf{TA:} Functions as a web server running within a TEE. %
    \item \textbf{Verifier:} Verifies the evidence on behalf of the users.
    \item \textbf{CA:} Issues the service certificates.
    \item \textbf{CT Logs:} Logs all service certificates.
    \item \textbf{CT Monitor:} Collects \& monitors certificates on CT Logs.
\end{itemize}

Users are expected to trust the TEE, CA, CT Logs, and CT Monitor while the service and Verifier are assumed to be malicious.
    A malicious service and Verifier's goal is to impersonate a TA and steal the users' confidential information or forge messages.

Note that we assume that the service and Verifier do not collude with each other.
    This concept is not novel; in fact, numerous systems proposed in the literature assume that entities are operated by organizations with different interests~\cite{ObiliousHTTP,ObiliousDNS,ObiliousDNSoverHTTPS}. %
We also assume that users follow general security practices, including checking the domain they are visiting and that all communication channels are protected via TLS.
    We acknowledge that having users check the domain is a challenge in itself, as shown by prior work~\cite{albakry2020url,reynolds2020measuring}.
    This is considered orthogonal to our work and we refer to tools proposed in the literature to aid users when checking the domain~\cite{althobaiti2018faheem,canova2015learn,kumaraguru2010teaching}.
Finally, we consider all side-channel and physical attacks against TEEs orthogonal to this work, and DoS (Deny of Service) attacks and malware on the user's device to be out of scope.

The threat model yields the following security requirements:
\begin{itemize}
    \item \textbf{Confidentiality.}
        Malicious service or Verifier cannot steal the data communicated between users and TA.
     \item \textbf{Integrity.}
        Malicious service or Verifier cannot tamper with users' requests or TA's responses.
\end{itemize}

We also set \textbf{Compatibility} as a non-security goal; the system must be deployable utilizing current off-the-shelf web mechanisms without the need of additional user-side software.

\section{Design}
\label{sec:design}

In this section, we first lay out strawman solutions and their limitations.
We then discuss the design of \sys and the challenges we faced during the design process.

\subsection{Strawman Solution}
One straightforward way of integrating RA into the web ecosystem is by utilizing client-side software, e.g., browser extensions.
    As per Section~\ref{sec:intro}, this approach is not ideal as users must install and configure software, which is against our compatibility requirement.

Another method is to extend the TLS protocol so that it incorporates RA.
    Although this approach is promising, it is currently undergoing active discussion, and therefore, unfortunately, does not meet our compatibility requirement.

The final approach involves utilizing a proxy server operated by a third party, which performs RA and key negotiation on behalf of the users.
    Because the proxy server plays a critical role in exchanging cryptographic artifacts that secures communication between the user and the TA, the proxy must be fully trusted.
    This does not fulfill our threat model where the verifier (i.e., the proxy) can be malicious.

\subsection{Conceptual Design}
Motivated by the limitations listed above, we construct \sys to avoid the discussed drawbacks.
\sys is inspired by the background check model of RATS~\cite{RATSRFC} (see Appendix~\ref{appendix:background:rats} and Figure~\ref{fig:background-check-model}).
    When contacted by the user (i.e., relying party), the TA (i.e., attester) provides the user with attestation evidence, which cryptographically attests to the genuineness of the TEE as well as the code running within the TA.
    This evidence is forwarded to the Verifier, which, on behalf of the user, checks whether it matches the expected value and then sends the result to the user.
    According to the verification result, the user decides whether to continue establishing a connection and using the service provided by the TA.

\subsection{Challenges \& Solutions} 
\label{sec:design:challenges_solution}
In realizing the design shown above, we identified the following challenges.
    Most of these challenges arise when applying RATS, an abstract protocol that encompasses different use cases, to the web context.

\subsubsection{Delivering information to users}
\leavevmode \\ \textbf{Challenge.}
During RA, users must check several parameters, including the validity of the TA itself and the code running within it.
    However, it is not clear how users obtain such parameters in a reliable and trustworthy manner.
The most straightforward method is delivering verification proofs and parameters via mediums that are accessible to the users.
    Indeed, many RA systems consider the use of X.509 extensions, including RATS~\cite{RATSRFC} and SGX-RA-TLS~\cite{knauth2018sgxratls}.
    However, even if users obtain such parameters and necessary cryptographic proofs, they still must verify them, which requires additional tools and steps, thereby leading to a loss of compatibility.

\leavevmode \\ \textbf{Solution.}
In \sys, the Verifier displays the verification outcome and TA information to users through its website.
    To enable this, the Verifier's website appears as a pop-up (e.g., in a new window or a new browser tab), where users can view the TA's information, including its parameters and the corresponding verification result.
    After confirming the validity, users can close the pop-up and resume interacting with the TA's website.
However, since the Verifier is considered untrusted, it may attempt to deceive users by displaying falsified verification results, thereby framing a legitimate TA.
    Although this scenario is beyond the scope of our threat model, we explore ways in which \sys can mitigate such framing attacks in Section~\ref{sec:further-enhancing-security} \& Appendix~\ref{sec:framing_attacks}.

\subsubsection{Continuous TA domain monitoring}
\label{sec:ta_monitoring}
\leavevmode \\ \textbf{Challenge.}
As users rely on the Verifier to verify TAs, it is crucial that the Verifier can distinguish between a legitimate and malicious TA.
    This is achieved by requiring the TA operator to hand attestation evidence and the code running within the TA to the Verifier\footnote{It is clear that this only works if we assume that the TA operator is willing to provide the source code to their TA, which is a drawback. However, this is a common issue regarding RA in TEEs.}.

The TA operator is also required to cryptographically bind the TA's public key and the domain name of the TA to the quote.
    However, \sys is susceptible to \emph{domain impersonation attacks}, where a malicious TA operator can change the IP address that is tied to the registered TA domain name in an attempt to direct users to a different, potentially malicious machine.

\leavevmode \\ \textbf{Solution.}
Domain impersonation attacks occur when malicious TA operators attempt to update the public key that is coupled to the TA's domain.
    With this observation, we designed \sys so that it leverages CT~\cite{CT1RFC, CT2RFC} to monitor for any changes in the public key that is tied to registered TA domains.
    By periodically querying a CT monitor, the Verifier collects every certificate issued by a CA under a certain domain, allowing it to detect any changes that were made to the public key that is registered under that domain.
    Another benefit of utilizing CT is that it is widely adopted by most modern, popular web browsers (see Appendix~\ref{appendix:browser-support} for details), making it ideal for the deployability aspect of \sys.

Note that CT logs are allowed to delay the logging of certificates up to a predefined Maximum Merge Delay (MMD, usually 24 hours~\cite{HowCTWork93:online}).
    Therefore, the downside of utilizing CT logs in the proposed system is that it enables certificates that impersonate TAs to be left undetected during the MMD.
    Moreover, CT monitors also introduce service delays~\cite{sun_certificate_2024}, which can further delay the detection.
    Although \sys does not actively mitigate this issue, Appendix~\ref{sec:collab-with-3rd-party} discusses potential solutions to this issue.

\subsection{The \sys Protocol}
\label{sec:protocol}

\begin{figure}[t]
  \centering
  \includegraphics[width=90mm]{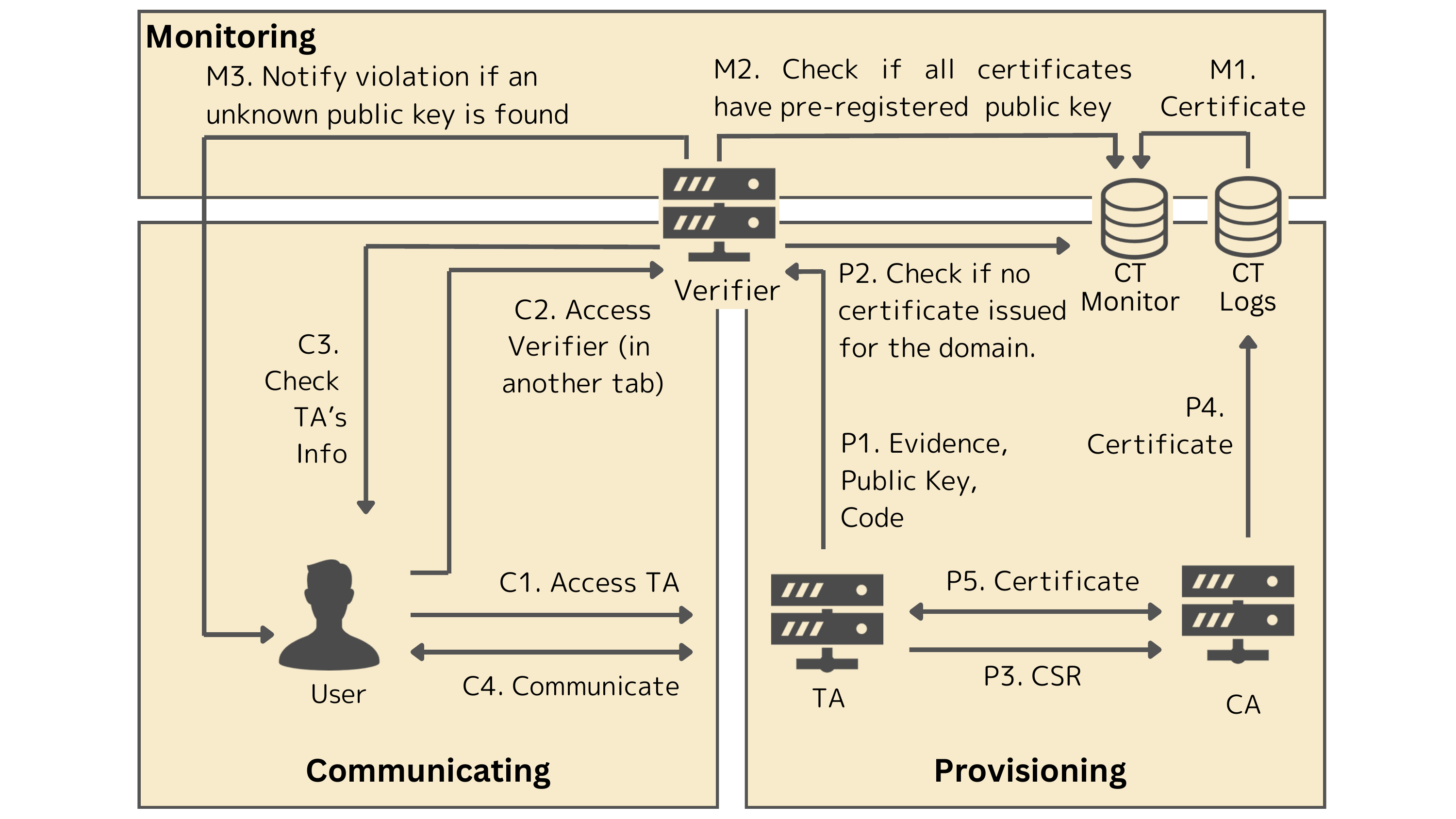}
  \vspace{-5mm}
  \caption{Overview of the \sys protocol.}
  \label{fig:attestation}
\end{figure}

{\sys} consists of the following three phases: Provisioning, Communicating, and Monitoring.
This section describes the details of each phase.
    Table~\ref{tab:notation} shows the notations used in the section.
    Figure~\ref{fig:attestation} shows the overview of \sys.

\begin{table}
    \centering
    \footnotesize
    \caption{Notation \& Definitions}
    \label{tab:notation} 
    \begin{tabular}{ll}
        \toprule
        Notation & Description \\
        \midrule
        $\object$ & Object (e.g., $\domain, \pk, \sk$) \\
        $\entity$ & Entity (i.e., $\user, \service, \ta, \verifier, \ctmonitor, \ctlog$) \\
        $\object.\object'$ & $\object$ of $\object'$ \\
        $\mlist_\object$ & List of $\object$  \\
        $\code$ & TA's code \\
        $\domain$ & TA's domain \\
        $\pk$ &  TA's public key \\
        $\sk$ &  TA's secret key (protected by TEE) \\
        $\evidence$ & Evidence (i.e., proof of remote attestation) bound with $\pk$ \\
        $\cert$ & Certificate of the TA \\
        $\precert$ & Precertificate of the TA \\
        $\sct$ & Timestamp by the CT Logs \\
        $\activate$ & Activation flag for TA  \\
        $\violation$ & Violation log for TA  \\
        $\code$ & The source code of TA \\
        $\rv$ & The TA identity (computed from $\code$), \\ & i.e., Reference Value in RATS (e.g., MRENCLAVE) \\
        $\tainfo$ & TA information, i.e.,  $(\domain, \rv, \code, \evidence)$ \\
        \bottomrule
    \end{tabular}
\end{table}

\subsubsection{Provisioning}
\begin{figure}
    \centering
    \input{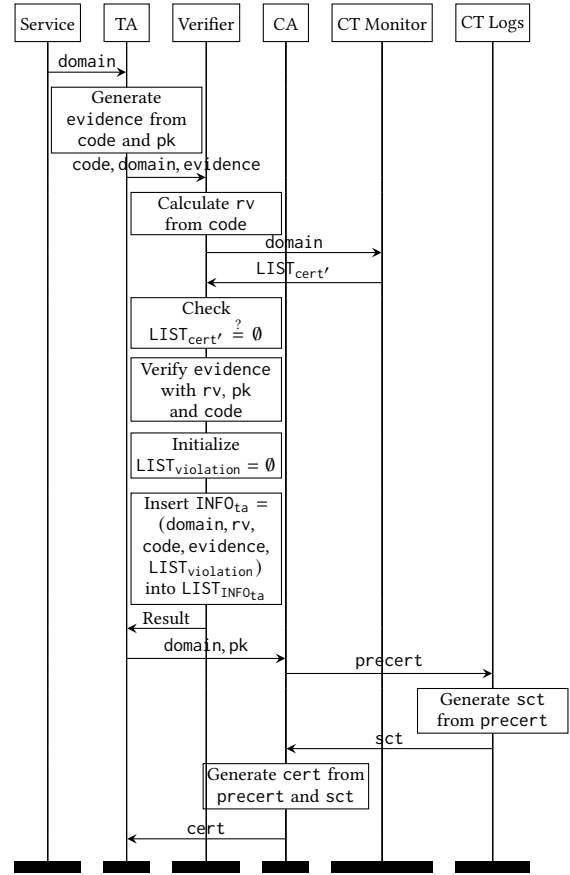}
    \caption{Provisioning Phase of \sys.}
    \label{fig:behavior-provisoning}
\end{figure}

During the provisioning phase, the TA initializes and sends its parameters to the Verifier, and the Verifier checks and stores them as shown in Figure~\ref{fig:behavior-provisoning}.

\subsubsection{Communicating}
\begin{figure}[t]
    \centering
    \input{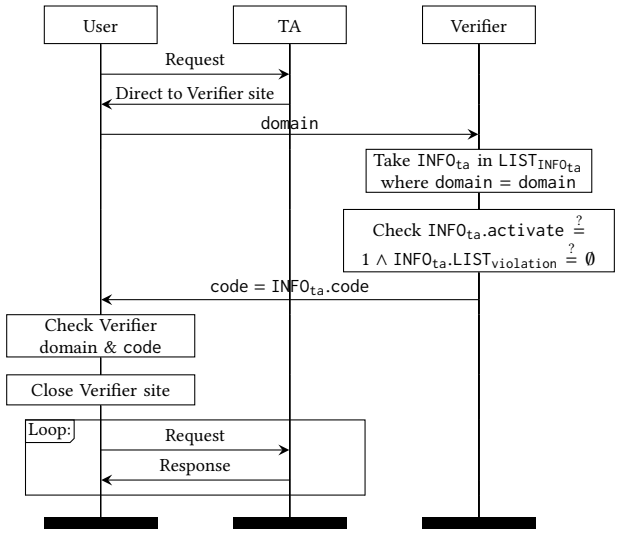}
    \caption{Communication Phase of \sys. }
    \label{fig:behavior-communication}
\end{figure}

The communicating phase consists of the process depicted in Figure~\ref{fig:behavior-communication}.
    In this phase, users first check the TA authenticity verification status on the Verifier's website shown in the pop-up, and after confirming the verification status, interact with the TA.
    The communication channel between the user and TA is secured through $\cert$.

\subsubsection{Monitoring}
\label{sec:protocol:monitoring}

\begin{figure}[t]
    \centering
    \input{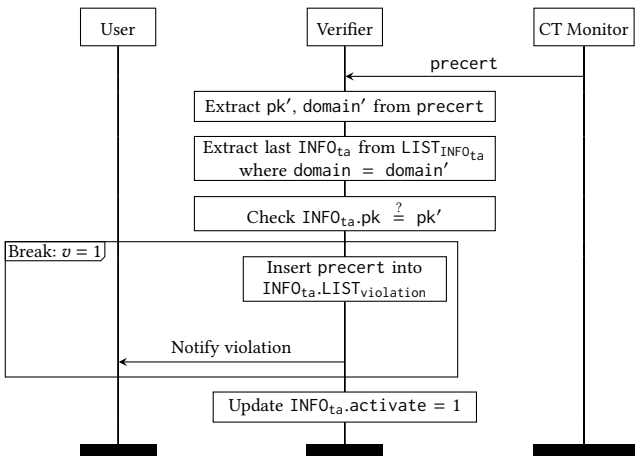}
    \caption{Monitoring Phase of \sys. $\precert$ is obtained from the CT log via the CT monitor.}
    \label{fig:behavior-monitoring}
\end{figure}

During the monitoring phase, the Verifier continuously scans the CT log for any signs of malicious behavior by the TA, as illustrated in Figure~\ref{fig:behavior-monitoring}.
    This involves the Verifier regularly contacting the CT monitor to check for newly issued certificates related to a specific domain and verifying whether the public key embedded in the new certificate is recognized.

If the Verifier detects any malicious activities by the service, it promptly notifies users, enabling them to cease using the TA in a timely manner.
    Consequently, user notification is a mandatory component of \sys.

\section{Implementation}

\begin{figure}[t]
  \centering
  \includegraphics[width=85mm]{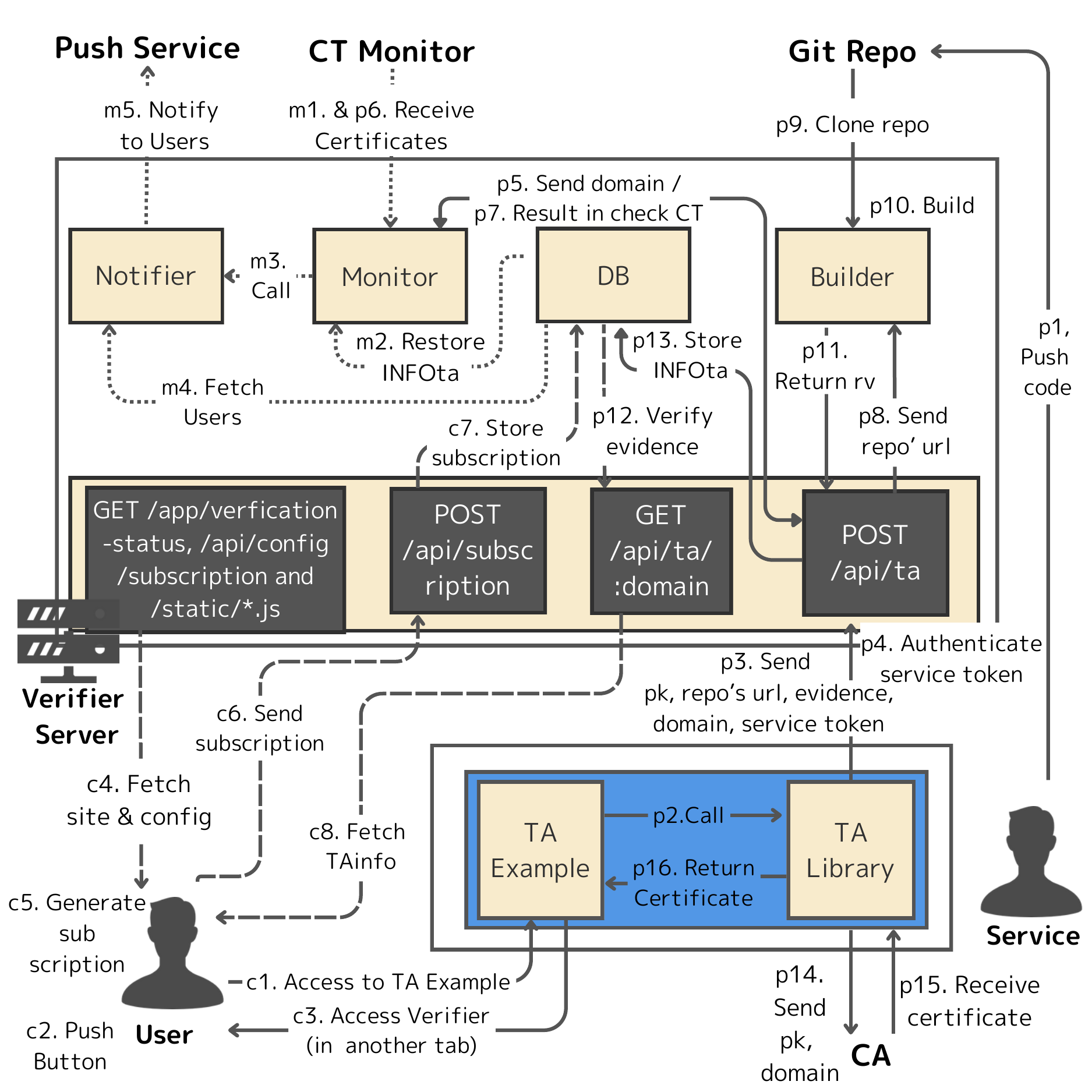}
  \caption{The flow and components in {\sys}. Each action is labeled with a letter denoting the corresponding phase, followed by a number indicating its sequence within the flow (e.g., \texttt{m4} represents the 4th action in the Monitoring phase). }
  \label{fig:impl}
  \vspace{-4mm}
\end{figure}

Based on the design presented in the previous section, we implemented the necessary components of \sys.
    Figure~\ref{fig:impl} shows the overall image of the implemented components and their interactions.

We developed \sys with Intel SGX~\cite{anati2013innovative} as the TEE and deployed it on Microsoft Azure.
    For ease of deployment, we leveraged Azure Data Center Attestation Primitives (DCAP)~\cite{AzureDCAP:online} as an attestation server. 
    All of our implementations are written in the Go language~\cite{GoLang:online}, chosen for its security, performance, ease of use, and the availability of extensive libraries.
    As for cryptographic primitives, we utilized RSA with 2048-bit keys for asymmetric encryption and SHA-256 as the secure cryptographic hash function. 
    It is important to note that \sys is not limited to a certain cryptographic primitive.

\subsection{Trusted Application}

The TA consists of a helper TA library and an example TA server.
    For the implementation, we adopt EGo~\cite{Ego:online} which is a Golang framework for building Intel SGX applications.
    In this section, we describe the two components in more detail.

\subsubsection{TA Library}

The TA library is designed to support any arbitrary TA that is built by EGo to attest itself utilizing the \sys protocol.
    Specifically, it is responsible for providing the following two functionalities: (1) Registering the TA to the Verifier, and (2) Requesting TLS certificates from the CA.
    To issue the certificate, we leverage Let's Encrypt~\cite{LetsEncrypt:online}. %

When invoked, the library generates an RSA key pair and produces a quote that cryptographically binds the TA code to the public key, if not done already.
    Subsequently, the library registers the TA to the Verifier by sending the TA's public key, evidence, domain, and service authentication token to the provisioning API defined in Section~\ref{sec:impl:verifier:api}.

Once the registration is finished, the library utilizes the ACME~\cite{ACMERFC} protocol to request a certificate from a CA.
    This certificate is handed to the TA (in our case, the example TA server) so that it can establish secure communication channels with its users.

\subsubsection{Example TA Server} \label{sec:impl:ta:server}

The example TA server is programmed to return simply ``hello'' when contacted by a user.
    To adhere to the \sys protocol, the TA program is integrated with the TA library we described earlier.

A challenge we encountered during the implementation was that users cannot be automatically shown the Verification Status Page, and must deliberately choose to visit it.
    Since visiting this page is crucial, we must require the TA server page to include a button that opens the status page as a pop-up.
        This is implemented using the \textit{window.open()} function.
    This also has the added benefit of allowing users to check the validity of the TA at any given time.
Naturally, a malicious TA would not include this button to evade detection.
    We discuss this issue and how \sys prevents it in Section~\ref{sec:security-analysis}.

\subsection{Verifier Server}
\label{sec:impl:verifier}

The Verifier server is comprised of five components: APIs, Verification Status Page, Builder, Monitors, and Notifier.
    We implemented the APIs and web pages utilizing the Echo framework~\cite{Echo:online}, and SQLite~\cite{SQLiteHo96:online} and the ent ORM library~\cite{Ent:online} for data storage and retrieval.

\subsubsection{Core APIs}
\label{sec:impl:verifier:api}

The Verifier offers various APIs, including registration of services and TAs, as well as provisioning of TA information.
    To enhance clarity and understandability, we designed these APIs according to the RESTful~\cite{RESTAPI} model.
    The Verifier primarily exposes the following APIs (additional APIs are provided in Appendix~\ref{appendix:impl-details:apis}):

\begin{itemize}
    \item \textbf{POST /api/ta}\footnote{When calling this API, the service is required to authenticate itself to the Verifier via service credentials (See Section~\ref{sec:further-enhancing-security} \& Appendix~\ref{sec:hold-responsibility} for more details).}:
        This API provisions a TA by downloading code from a specified repository URL, verifying the service token and quote, checking for existing certificates, and storing TA information.
            The Builder component handles code downloading and building (see Section~\ref{sec:builder}).

    \item \textbf{GET /api/ta/:domain\footnote{\textbf{``:domain''} is a URL parameter expression. In this case, it refers to the TA domain.}}:
        This API retrieves TA information required for generating the Verification Status Page (see Section~\ref{sec:user-pages}) by searching the database using the provided domain parameter in the URL.

\end{itemize}

\subsubsection{Verification Status Page}
\label{sec:user-pages} 

When visiting the TA, the TA instructs the user to visit the \textit{GET /app/verification-status} API, where the Verifier sends the Verification Status Page and accompanying front-end JavaScript code.
This code enables the user's browser to verify the TA's validity and establish user notifications.
In this section, we provide a detailed explanation of this process.

\textit{Checking the validity of TA: }
Once loaded on the user's browser, the Verification Status Page uses its JavaScript code to check the TA's validity. %
    Specifically, the code retrieves the TA's domain and fetches the TA information from \textit{GET /api/ta/:domain}, the most important piece of information being the \emph{valid flag}.
    If this flag is disabled, the website communicates this to the user by displaying an error message.

The challenge in this process is securely passing the TA domain information to the Verifier.
    A straightforward approach would be to use query parameters, although this method is vulnerable to URL tampering attacks, which can lead to TA impersonation attacks.

In \sys, we overcome this issue by utilizing \emph{HTTP Referer Headers}~\cite{MDNRefererH75:online}, which enables the browser to know who referred the user to the Verifier via the HTTP header.
    This allows the user's browser, as the referred entity, to understand the domain from which it was referred from, thereby preventing malicious TAs from tampering with the TA domain name and mitigating URL tampering attacks.

\textit{Establishing User Notifications: }
Recall that user notifications are a vital component of \sys, enabling users to stay informed about malicious TAs (see Section~\ref{sec:protocol:monitoring}).
    The Verification Status Page facilitates the setup of these notifications.
    To initiate this process, users first fetch global parameters (e.g., Verifier's public key for  Notification~\cite{VAPIDRFC}) via the \textit{GET /api/config/subscription} API.
    With this information, users generate a unique ``subscription'' token, which is a public key created by the user-side JavaScript code~\cite{rfc8291}.
    This token is then sent to the Verifier via the \textit{POST /api/subscription} API (shown in Section~\ref{sec:notifier}), registering the user for notifications.

Note that browsers require a script to receive notifications, which is implemented as a Service Worker.
    This program runs continuously on the user's device, allowing users to receive notifications even after closing their browser.
    Although using Service Workers may seem to hinder \sys's compatibility aspect due to the required installation, these scripts are automatically installed without user interaction.
    Moreover, Service Workers are strongly isolated by the browser's isolation technology and have limited accessible APIs, mitigating security concerns.

\subsubsection{Builder}
\label{sec:builder}

The Builder component is responsible for calculating the $\rv$ for quote verification.
    This process involves cloning the git repository, building the code, and then calculating the $\rv$ value.
    Finally, the Builder checks whether the calculated $\rv$ value matches the one included in the $\evidence$.
    In this implementation, the $\rv$ value is calculated using the \texttt{ego uniqueID} command~\cite{uniqueid:online}, which generates the hash of the program executable.

\subsubsection{Monitor}
\label{sec:monitor}

As previously noted, the monitor is employed in \sys to detect any changes to the TA certificate.
    For the implementation, we chose to utilize \texttt{crt.sh}~\cite{crtshCer92:online}, a third-party CT monitor.
    This service provides an ATOM feed that lists all certificates for a given domain, allowing us to access logged certificates for a specific domain via a URL. %

Note that CT monitors are not perfect; in fact, a recent study~\cite{sun_certificate_2024} revealed that some CT monitors fail to collect every certificate from the CT log.
    We chose \texttt{crt.sh} because \cite{sun_certificate_2024} reported that it was the only monitor to report no certificate omissions during the study period.

\subsubsection{Notifier}
\label{sec:notifier}

The Notifier is a component that notifies users about updates or violations of the TA. 
    In our implementation, we utilize the W3C Push API~\cite{PushAPI57:online} and the Notification API~\cite{Notifica41:online} mechanisms to send push notifications via browser vendors' push service.
    It is important to note that \sys is not limited to a certain notification mechanism and can be easily replaced with other mechanisms.
    The notification is delivered to the designated destination by using the subscriptions provided by users.
    Users interact with the APIs shown below to set up such subscriptions:

    \begin{itemize}
    \item \textbf{POST /api/subscription}:
        This API receives and stores users' subscription information, such as push notification details (also see Section~\ref{sec:user-pages} for details).
    
    \item \textbf{GET /api/config/subscription}: 
        This API provides users with global parameters, such as the public key of the TA, required to generate subscription tokens. 
\end{itemize}

\section{Security Analysis}
\label{sec:security-analysis}
This section provides the security analysis of \sys based on the threat model defined in Section~\ref{sec:system}.

\taggedpara{Machine In the Middle Attack (MITM)}
In this attack, the service attempts to intercept, eavesdrop on, and tamper with communications between the TA and the user.
However, this attack is prevented by securing the communication channel between the TA and the user using TLS.
    Note that services may generate false certificates and impersonate a TA, but the Verifier can detect this (see next).

\sys can prevent MITM attack even in the case where the Verifier is malicious, thanks to the secure channel that is established between the TA and the user.

\taggedpara{Domain impersonation attack}
As mentioned in Section~\ref{sec:design:challenges_solution}, the service may assign the TA's domain to another machine, thereby launching a Domain Impersonation Attack.
    This allows the service to impersonate the TA and intercept the communication from the user, thus violating the confidentiality requirement.

RA-WEBs can prevent this attack by requiring the Verifier to monitor the certificates issued to the TA's domain by utilizing the CT mechanism.
    If a certificate containing a public key other than the TA's is issued to the TA domain, the Verifier can detect it and notify the user.

\taggedpara{Re-registering different types of TA}
A malicious service operator may update and re-register a different (potentially malicious) TA in place of the original one after the user successfully conducted RA.
    This is an issue because \sys allows users to skip verification from the second time onward to improve usability.
    To prevent this, the Verifier notifies users that a TA has been re-registered and must once again go through the verification process.

\taggedpara{Registering malicious TAs}
A service operator may deploy a malicious TA that leaks sensitive data (e.g., the TA private key), or act as a proxy server to a non-protected service.
    The functionality of the TA could also be tampered with so that it does not direct users to the Verifier even when it must do so.

\sys prevents this by requiring the user to verify that the TA reference value is what they trust.
    If the TA source code is publicly available, it is also recommended to check for any malicious code.
        We discuss how we aim to support users without any technical knowledge in Section~\ref{sec:usability} \& Appendix~\ref{sec:code-auditing}.
Once a user finds a service deploying a malicious TA, they must take action, which is not straightforward.
    We refer the reader to Section~\ref{sec:further-enhancing-security} \& Appendix~\ref{sec:hold-responsibility} for more details.

\taggedpara{Services attempting evasion}
Some malicious services may not direct users to the Verifier in an attempt to prevent users from verifying the TA's authenticity.
    This is not meaningful, because users can easily detect that the service is not directing them to a Verifier, which is a tell-tale sign that it is acting maliciously.

\taggedpara{Services impersonating Verifier}
A malicious service may impersonate a Verifier, directing a user to a fake verification site.
    This enables the service to deceive the user into believing the TA is legitimate by displaying fabricated verification results.
    \sys mitigates this risk by ensuring that users verify the domain they are visiting is indeed the trusted Verifier.

\taggedpara{Non-mainstream browsers}
It is possible for malicious services to attack users that utilize browsers that are not supported in \sys.
    For example, the service may issue certificates to CT logs that are not recognized by the Verifier but are so by certain browsers.
    Furthermore, some browsers may not be compatible with \sys.
        For instance, users using browsers that lack support for push notifications would not detect domain impersonation attacks.

Users utilizing these non-mainstream browsers can be detected and notified of the risks.
    Specifically, the Verifier can check the User-Agent header, and display information to the User if the browser is not supported.
    We also discuss how we can utilize a special kind of Verifier to support such users (see Section~\ref{sec:openness-generality} \& Appendix~\ref{sec:non-mainstream-browsers}).

\taggedpara{Utilizing previously issued certificates} 
A malicious service may withhold information from the Verifier about a previously issued certificate for the domain, which was obtained during the provisioning phase. 
    This omission enables the service to launch domain impersonation attacks, as both the previously issued certificate and the new certificate remain valid.

To prevent this attack, \sys mandates that the Verifier checks for any valid certificates associated with the TA domain during the provisioning phase. 
    If a valid certificate is found, the Verifier terminates the TA registration process.

\taggedpara{Impersonation during the CT delay}
As mentioned previously, there is a delay when the CT records a certificate.
    During this time, a service can impersonate a TA without being detected.
Although we cannot fully prevent this attack, \sys can reduce the risk by limiting the delay, as shown in Section~\ref{sec:further-enhancing-security} \& Appendix~\ref{sec:collab-with-3rd-party}.
    Moreover, holding services responsible for their actions can further reduce the risk of misbehaving services (see Section~\ref{sec:further-enhancing-security} \& Appendix~\ref{sec:hold-responsibility}).

\taggedpara{Directing users to malicious TA}
A malicious Verifier may attempt to obtain users' data by redirecting them to a compromised TA operated by the Verifier.
    However, this attack is prevented in \sys because the TA is the entity that directs users to a Verifier, not vice versa.

Alternatively, a malicious Verifier might redirect users to a fraudulent website that impersonates a TA after they visit the Verifier's website.
    Fortunately, users can detect this attack because the Verifier's website they visited no longer displays the Verification Status Page (described in Section~\ref{sec:user-pages}).
    Moreover, we can further prevent such an attack by mandating that Verifiers operate within a TEE, as discussed in Appendix~\ref{appendix:ta-verifier}. \\

\taggedpara{Summary}
Overall, we argue that {\sys} meets the security requirements defined in Section~\ref{sec:system}. %

\section{Discussion} 
\label{sec:discussion}

In this section, we provide a summary of additional discussion points which were not covered in the main body.
The details of each discussion can be found in Appendix~\ref{appendix:discussion-details}.

\subsection{Further enhancing \sys security}
\label{sec:further-enhancing-security}
The security of \sys can be enhanced in several ways. 
Firstly, the risk of domain impersonation can be mitigated by minimizing CT delay.
    Although the CT delay is smaller than the time it takes to discover backdoors planted in Open Source Software~\cite{xzbackdoor:online}, it is still beneficial for users if we can minimize the delay introduced by CT.
    This can be achieved through various means, such as deploying a CT monitor with a shorter delay, running our own CT monitor, or leveraging CT logs with minimal delays.

Additionally, \sys can hold services accountable for their actions with the help from the Verifier.
    This can be accomplished by requiring services to authenticate themselves to the Verifier and sign a contract during the provisioning phase.

To further fortify the security of the system, \sys can implement several modifications to the Verifier.
    For instance, to protect users even in cases where Verifiers and services collude, \sys can introduce multiple Verifiers that monitor each other for any malicious behavior.
    Moreover, running Verifiers within a TEE can provide an additional layer of security, enabling entities to monitor and cryptographically verify the Verifier's behavior, as well as benefit from features such as non-repudiable proofs and enhanced user privacy.

Furthermore, \sys can collaborate with third-party entities to revoke malicious TAs and further reduce CT delay.
    Users who prioritize security over compatibility can also install additional user-side software, such as custom SSL libraries and browser extensions, to provide an extra layer of protection.

Lastly, we discuss attacks that were not covered in Section~\ref{sec:security-analysis} due to the adversary models that were not considered in Section~\ref{sec:system}, including framing attacks and detecting TAs running on vulnerable CPUs.

\subsection{Usability analysis}
\label{sec:usability}
Usability is a crucial aspect of \sys.
    We focus on two key discussion points: UI fatigue and source code checking.
We argue that the UI of \sys does not impose a significant burden on users, thereby mitigating the risk of UI fatigue.
    Moreover, the load of source code checking can be alleviated by leveraging the collective efforts of security-conscious users, who can review TA code on behalf of the majority of users.

\subsection{Openness analysis}
\label{sec:openness-generality}
As \sys operates in the web environment, openness is a vital aspect to consider.
\sys has been designed and implemented using standard cryptographic protocols and features that are widely adopted by modern browsers.
    To ensure openness, we evaluated its adoption on 10 browsers.

To further enhance openness, we explore the possibility of extending \sys to incorporate alternative notification methods, such as email.
    Furthermore, we examine how non-mainstream browsers can be integrated into the \sys ecosystem.
    We demonstrate that this can be achieved by allowing users to deploy their own Verifiers.

Furthermore, there are standards being developed that focus on delivering attestation evidences embedded in certificates~\cite{ietf-rats-pkix-evidence-00, X509RA}.
    Once these standards become widely available and browsers start to integrate this feature, \sys can benefit from this in the context of openness, as it enables anyone to audit the TA.
    This is because RA proofs can be publicly viewed via CT logs.

\subsection{Load evaluation}
\label{sec:load}
One of the non-security objectives outlined in Section~\ref{sec:system} is to reduce the load on the Verifier.
In this context, we examine the load experienced by the Verifier across five distinct aspects: API access, CT monitoring, user notification, code building, and Verifier-side storage.
    Our experimental results demonstrate that API access and Verifier-side storage do not incur significant loads.
    In contrast, the loads associated with CT monitoring, user notification, and code building can be effectively distributed across the entire \sys operation, thereby mitigating their impact.

\subsection{Deployment considerations and incentives}
\label{sec:deployability}
Understanding the considerations and incentives for deploying \sys is essential for enhancing its immediate deployability.
Verifiers can be operated by various organizations with diverse incentives.
    For example, TEE hardware vendors may operate Verifiers to increase the appeal of their hardware.
    It is also noteworthy that having multiple Verifiers in the system is beneficial for the overall \sys ecosystem.
Additionally, services can benefit from integrating \sys into their web services, as it enables them to demonstrate their commitment to usability, security, and privacy.

\section{Related Work}

\taggedpara{Continuum Verifier~\cite{Continuum}}
The attestation infrastructure employed by Continuum, a platform offering privacy-preserving conversational LLMs that leverages TEEs to protect users' LLM prompts, is the closest existing service to our system.
    In Continuum's setup, a proxy server acts as a verification service, enabling users to delegate verification tasks and generating a secret key to encrypt user LLM prompts upon successful verification.
    This approach offers benefits, such as no browser modifications and immunity to CT log delays, but has a significant security limitation: the proxy server controller can decrypt and access user-submitted LLM prompts.
    To mitigate this, Continuum offers users the option to deploy their own proxy server, but this requires installation and deployment, violating our compatibility requirement.
    Even if a third party operates the proxy server, confidentiality risks remain, unlike \sys, which provides a more secure solution, mitigating the risk of content exposure to malicious Verifiers, as discussed in Section~\ref{sec:security-analysis}.

\taggedpara{SGX-RA-TLS~\cite{knauth2018sgxratls}}
SGX-RA-TLS integrates RA into the TLS connection establishment procedure.
    This integration is achieved by embedding the attestation evidence into the X.509 extension or existing X.509 fields, such as the \texttt{CommonName} field.
    While this approach offers numerous benefits, including no required changes to the protocol itself or its implementation, it does require the client to download additional software to facilitate the special TLS handshake procedure, which, unfortunately, does not meet our compatibility requirement.

\taggedpara{Integrating Attestation into TLS \& DTLS~\cite{fossati-tls-attestation-07}}
An IETF working group is proposing a series of extensions to the TLS 1.3 protocol standard to integrate attestation capabilities.
    The primary advantage of this protocol is that it enables attestation to occur during the TLS handshake, rendering the attestation process completely transparent to the user, much like the TLS protocol itself.
    However, modifications to such a low-level protocol often require a lengthy standardization process and development process which unfortunately does not align with our compatibility requirement.

\taggedpara{Remote ATtestation procedureS~\cite{RATSRFC}}  
The Remote ATtestation procedureS (RATS) standard defines the entities and their interactions necessary for securely conducting RA.
    The primary objective of RATS is to provide a model that is agnostic to processor architectures, protocols, and data exchanges, which has led to the specification of an abstract, high-level model.
    Although the design of \sys was inspired by the RATS background model, \sys focuses on highlighting the challenges we encountered when adapting the model to the Web context and exploring its feasibility within the constraints of compatibility and openness.

\taggedpara{Confidential Computing Transparency~\cite{kocaoğullar2024confidentialcomputingtransparency}}
Motivated by the fact that RA alone cannot guarantee the absence of backdoors and vulnerabilities, Confidential Computing Transparency (CCT) proposes a framework that ensures the verifiability of a TA's executable binary by leveraging CT, essentially introducing Binary Transparency to the confidential computing ecosystem.
    By doing so, CCT empowers users and interested parties to audit the TA code, enabling them to detect potential issues in a timely manner.
While CCT does not focus on integrating RA into the web ecosystem, we believe it complements \sys, thereby further enhancing its security features.

\section{Conclusion \& Future Work}

This work proposes \sys, a novel RA protocol designed for high compatibility with the current web ecosystem by utilizing commonly available web mechanisms, which also enables users to verify the validity of the RA verification result without installing any software on their side.
Our proof-of-concept implementation shows that the proposed method is immediately deployable, and through extensive security and load evaluation, we demonstrate that \sys is secure against various attacks and demonstrates minimal load.
As for future work, we plan to investigate possibilities of integrating the ideas discussed in Appendix~\ref{sec:discussion}, expand \sys implementation to support different types of TEEs, and deploy our system in the real world.

\bibliographystyle{templates/ACM-Reference-Format}
\bibliography{citations}

\newpage

\appendix
\section{Background}
\subsection{Trusted Execution Environment}

Trusted Execution Environment (TEE) is a security primitive that protects data and code from adversaries, including privileged software and hardware owners~\cite{confidential2020confidential}.
    A typical TEE has the following functionalities:
    (a) Isolated Execution: provides an execution environment where data and code are isolated from all other software on the platform, including privileged system software (e.g., OS, hypervisor, BIOS); and
    (b) Remote Attestation (RA): provides a remote party with strong cryptographic assurance that the expected program is running within a valid TEE.

\subsection{Remote ATtestation ProcedureS} \label{appendix:background:rats}

\begin{figure}[t]
  \centering
  \includegraphics[trim=1cm 9cm 1cm 7cm,clip,width=75mm]{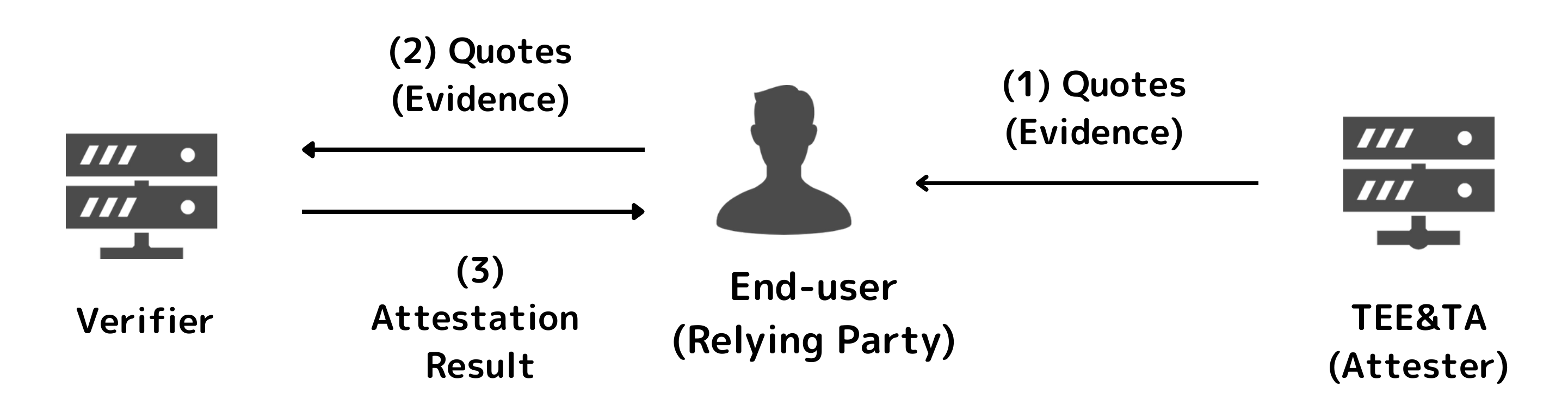} 
  \caption{Background check model of RATS~\cite{RATSRFC}. Parentheses represent terms used in RATS.}
  \vspace{-5mm}
  \label{fig:background-check-model}
\end{figure}

Remote ATesstation procedureS (RATS)~\cite{RATSRFC} is a standardized RA protocol defined at RFC 9334, which aims to define common terminology, roles, and procedures to provide interoperability across platforms as well as to support many RA use cases, facilitating openness.
    RATS provides two communication models, namely passport and background check.
        Here, we focus on the background check model (shown in Figure~\ref{fig:background-check-model}).
        In this model, RATS introduces a trusted third-party Verifier in addition to the Attester (e.g., TA) and Relying Party (e.g., User).
        The Verifier acts as a cushion between the Relying Party and the Attester, verifying Evidences on behalf of the Relying Party, which may not always have the capability or may not be compatible with the verification protocol.

\subsection{Certificate Transparency}
Certificate Transparency (CT) \cite{CT1RFC, CT2RFC}is a system designed to improve security in the Web PKI using cryptographically verifiable append-only logs.  Originally proposed by Google and standardized as RFC 6962~\cite{CT1RFC}, CT is currently in its second revision~\cite{CT2RFC} and is used by almost all participants in the Web PKI ecosystem.

CT forces certificate authorities to disclose issuance for all certificates in order to be trusted by CT-enforcing clients (such as Google Chrome).  A CT log issues a signed timestamp (SCT) based on a certificate's contents, which represents a promise to publish the certificate data publicly for anyone to download or audit within a specific timeframe, known as maximum merge delay (MMD).  Log monitors and auditors can download the contents of the log and inspect for discrepancies from policy or to help domain administrators monitor for certificate issuance using their domains.

\section{Implementation Details}
\subsection{Additional Endpoints}
\label{appendix:impl-details:apis}
In addition to the endpoints shown in Section~\ref{sec:impl:verifier:api}, Verifier has the following APIs:

\begin{itemize}
    \item \textbf{POST /api/service}: 
        This API is utilized during the service registration process.
            When invoked, the Verifier authenticates the Verifier owner's credentials, generates a new token and a unique identifier for the service, stores both the token and identifier, and sends the token to the service.

    \item \textbf{POST /api/notify}: 
        This API is for the verifier owner to notify some information to users. 
        For example, the service may tell users the plan of the temporary stop service for the verifier maintenance.

\end{itemize}

\subsection{Database}
\label{appendix:impl-details:dbs}

In Section~\ref{sec:impl:verifier}, we leveraged a database (i.e., SQLite) to store critical information such as TA information, subscription details, and violation records. 
    During the development of \sys using the database, we defined the following tables:

\begin{itemize} 
    \item \textbf{serivce:} This table has information of the service (Table~\ref{table:service-table}). 
    \item \textbf{ta\_codes:} This table stores the information regarding source code run by TAs (Table~\ref{table:ta_code-table}). 
    \item \textbf{ta\_servers:} This table contains the servers on which the TAs are running (Table~\ref{table:ta_servers-table}). 
    \item \textbf{ta\_violations:} This table logs violation associated with the TAs (Table~\ref{table:ta_violatios-table}). 
    \item \textbf{subscriptions:} This table holds data related to notification subscriptions (Table~\ref{table:subscription-table}). 
\end{itemize}

Notably, we designed the TA structure by separating it into \texttt{ta\_servers} and \texttt{ta\_codes}. 
It offers potential benefits for future scalability and efficiency. 
    For instance, the Verifier could reuse the same \texttt{ta\_codes} for multiple servers, thereby reducing the overall storage required. 
        Although this design has not yet been fully implemented, we plan to implement it in the future.
    Additionally, this modular design would allow the Verifier to revoke multiple TAs associated with the same malicious code simultaneously. 
        By simply toggling a disable flag, the Verifier can easily ban all affected TAs at once, streamlining the revocation process.
    
Another notable design choice is for the Verifier to store the monitor\_log\_id (i.e., Certificate ID from crt.sh) instead of the full precertificate. 
    This approach significantly reduces storage consumption, 
        shrinking the data stored from the size of the certificate itself to just a small numeric identifier.

\newpage

\begin{table}[!h]
    \centering
    \footnotesize
    \caption{The columns in \texttt{services}}
    \begin{tabular}{lcl}
        \toprule
        Name & Type & Description \\
        \midrule
        id & Integer & Primary key \\
        name & String & Service name (empty in current) \\
        token & String & Credential \\
        is\_active & Bool & The flag indicates that \\ & & service is enabled \\
        \bottomrule
    \end{tabular}
    \label{table:service-table}
\end{table}

\begin{table}[!h]
    \centering
    \footnotesize
    \caption{The columns in \texttt{ta\_codes}}
    \begin{tabular}{lcl}
        \toprule
        Name & Type & Description \\
        \midrule
        id & Integer & Primary key \\
        repository & String & URL of Git Repository \\
        commit\_id & String & Commit ID of Git Repository \\
        unique\_id & Bytes & MRENCLAVE in Intel SGX (i.e.,  $\rv$) \\
        is\_active & Bool & The flag indicates that code is enabled \\
        ta\_code\_service & Integer & The ID of owner, i.e., service \\
        \bottomrule
    \end{tabular}
    \label{table:ta_code-table}
\end{table}

\begin{table}[!h]
    \centering
    \footnotesize
    \caption{The columns in \texttt{ta\_servers}}
    \begin{tabular}{lcl}
        \toprule
        Name & Type & Description \\
        \midrule

        id & Integer & Primary key \\
        domain & String & $\domain$ of the TA \\
        public\_key & Bytes & $\pk$of the TA \\
        quote & String & $\evidence$ of the TA \\
        monitor\_log\_id & Integer & Certificate ID in crt.sh \\

        is\_active & Bool & The flag indicating that \\ & & TA has been activated \\
        ta\_server\_service & Integer & The ID of TA's code \\
        ta\_server\_code & Integer & The ID of TA's owner, i.e., service \\
        \bottomrule
    \end{tabular}
    \label{table:ta_servers-table}
\end{table}

\begin{table}[!h]
    \centering
    \footnotesize
    \caption{The columns in \texttt{subscriptions}}
    \begin{tabular}{lcl}
        \toprule
        Name & Type & Description \\
        \midrule

        id & Integer & Primary key \\
        endpoint & String & A part of subscription  \\
        p256dh & String & A part of subscription \\
        auth & String & A part of subscription \\
        subscription\_server & Integer & The ID of TA' server \\
        \bottomrule
    \end{tabular}
    \label{table:subscription-table}
\end{table}

\begin{table}[!h]
    \centering
    \footnotesize
    \caption{The columns in \texttt{ta\_violations}}
    \begin{tabular}{lcl}
        \toprule
        Name & Type & Description \\
        \midrule
        id & Integer & Primary key \\
        created\_at & Datetime & Date and Time \\ && when this column created  \\
        ta\_violation\_server & Integer & Violating TA \\
        ta\_violation\_service & Integer & Violating service \\
        \bottomrule
    \end{tabular}
    \label{table:ta_violatios-table}
\end{table}

\newpage
\subsection{Source Code of Example TA}

Listing~\ref{lst:example-ta} presents the source code of a sample TA that utilizes the \sys package, written in Golang.
    As shown, integrating \sys into a TA is a straightforward process.
    The software engineer simply needs to specify a path that points to the Verification Status Page via a pop-up button. %
    After the user closes the Verification Status Page, the TA operation resumes.

\lstset{style=gostyle}
\begin{lstlisting}[caption={Source Code of Example TA},label=lst:example-ta]
package main

import (
	"fmt"
	"net/http"

	"github.com/XXXX-1/ra_webs/ta"
)

const VERIFIER_PATH = "/app/verification-status/"

func main() {
	config, err := ta.DefaultConfig()
	if err != nil {
		panic(err)
	}

	ta, err := ta.InitTA(config)
	if err != nil {
		panic(err)
	}

	handler := func(w http.ResponseWriter, r *http.Request) {
		w.Header().Set("Content-Type", "text/html")

		fmt.Fprintln(w, "Hello from TA running on TEE :)<br/>")

		for _, v := range config.Verifiers {
			fmt.Fprintf(w, `<button onclick="window.open('%s');">Verifier Page (%s)</button><br/>`, v+VERIFIER_PATH, v)
		}
	}

	tlsConfig, err := ta.TLSConfig()
	if err != nil {
		panic(err)
	}

	server := http.Server{
		Addr:      ":443",
		Handler:   nil,
		TLSConfig: tlsConfig,
	}

	http.HandleFunc("/", handler)

	err = server.ListenAndServeTLS("", "")
	panic(err)
}
\end{lstlisting}

\section{Detailed Discussion} 
\label{appendix:discussion-details}

\subsection{Further enhancing \sys security}

\subsubsection{Security risk of CT delay}
\label{sec:security-risk-delay}

As previously mentioned, a service can impersonate a TA until the TA's certificate is detected by CT Monitors. 
    Previous research showed that the majority of CT monitors experience delays of several days (e.g., 6 days for 99.99 \% of certificates in crt.sh~\cite{sun_certificate_2024}).
We believe that such a delay is not a significant issue if the Verifier can hold the service accountable during this period (see Appendix~\ref{sec:hold-responsibility}). 
    Moreover, since it takes several days to months for a vulnerability to be found and mitigated on Open Source Software~\cite{xzbackdoor:online}, we argue that the delay introduced by CT monitors is negligible.

However, we also have options to mitigate or eliminate this security concern. 
    One option is to adopt a CT Monitor with a shorter delay.
        For example, \cite{sun_certificate_2024} reported that the delay of SSLMate Spotter is just 2 days, although this requires a paid plan.
    Alternatively, deploying our own CT Monitor could further reduce the delay to a few minutes~\cite{sun_certificate_2024}, which entirely mitigates the security risks posed by the delay. 
    Furthermore, leveraging CT logs that promise minimal delays, such as Sunlight~\cite{sunlight:online}, could further enhance security.
    However, it is important to note that there is a trade-off between minimizing CT monitoring delays 
        and the resource demands placed on the Verifier. 
        For example, an in-house CT Monitor would need to process approximately 10 million certificates per day, storing a total of 17 TB of data. 
        This highlights the operational overhead required to achieve shorter delays, which must be balanced against the security benefits.
    Finally, we can also eliminate the delay with the 3rd parties' help (See Appendix~\ref{sec:collab-with-3rd-party}).

\subsubsection{Pursuing Violation Responsibility}
\label{sec:hold-responsibility}

Some users may notice having interacted with a malicious TA after sending sensitive data, or the user may not have even been able to notice at all due to the CT log delay.
In such cases, users may attempt to hold service operators accountable with the help of the Verifier.
    One way of pursuing responsibility is to have the service sign a contract with the Verifier in advance, e.g., pose fines for violating terms and conditions.
Although some services may deceive about their fraudulent activities, since the Verifier possesses evidence such as certificates for domains and the $\rv$ of the TA, the Verifier can prove the wrongdoings of the service.
    To prove malicious activities with a higher degree of certainty, Verifiers should be mandated to run within a TEE (see Appendix~\ref{appendix:ta-verifier}) so that prosecutors are able to obtain non-repudiable proofs.

Note that for the above to function, services must provide an identifiable document (e.g., corporate authentication, proof of identity) and must be authenticated by the Verifier before provisioning.
    This is a limitation to this approach, as it only allows entities that have the capacity to register themselves as a corporation or a verified individual to launch TAs, which reduces the openness of \sys.

\subsubsection{Colluding Verifier and service}
\label{sec:distributed-verifier}
There may be instances where both the Verifier and Service could be operated by the same entity. 
    In particular, the Verifier may collude with a malicious service and produce genuine, valid proofs for malicious TAs.
Although this work assumes that either one of the Verifiers or Service is honest, \sys is capable of protecting users from such threat by introducing multiple, non-colluding Verifiers to the protocol.
    During the provisioning phase, a TA will be registered to multiple Verifiers, and users are directed to each Verifier during the communication phase.
    Under the assumption that users stop trusting all Verifiers whenever they are notified of the presence of a malicious Verifier, this model requires only a single legitimate Verifier to detect the presence of a malicious Verifier.
    There are two challenges to this approach.
    Firstly, latency during the communication phase would be larger compared to the case where we trust the Verifier.
        We see this as a necessary trade-off between security and usability.
    Secondly, malicious Verifiers can trivially DoS the \sys system by sending out falsified notifications.
        Instead of distrusting all Verifiers whenever there is a malicious Verifier, we can have the Verifiers themselves decide who is trustworthy. 
        To do this, we can have the Verifiers form a Byzantine Fault Tolerant (BFT)~\cite{bftlamport2019byzantine, bftcastro1999practical} system and reach a consensus on which Verifier is trustworthy.

\subsubsection{Running Verifiers within TEEs}
\label{appendix:ta-verifier}

Running Verifiers within a TEE yields several benefits:
    Firstly, these Verifiers will generate non-repudiable proofs, which can be used by a mediator (e.g., a court of justice) during its prosecution.
    Secondly, this approach can prevent Verifiers from tracking users' browsing history of TAs, for instance, by using cookies.
        This is possible because users can verify that the Verifier is executing the expected code, which explicitly excludes tracking their behavior.
        Although this is not part of our security requirement, it is a crucial feature for preserving user privacy.

However, a malicious Verifier operator can falsely claim that it is using a TEE or run a modified, malicious Verifier within a TEE.
    Therefore, it is crucial for external parties to verify the deployed Verifier.
    See Appendix~\ref{sec:distributed-verifier} for more discussion on how we achieve this.

\subsubsection{Securing \sys with 3rd Parties}
\label{sec:collab-with-3rd-party}

Although we have shown that \sys achieves the necessary security goals, we could further enhance its security by cooperating with entities, specifically Domain Registry (DRs; e.g., Google Domain) and CAs.
    Note that, this causes a trade-off between the burden posed to different entities and the achieved.
    In this section, we discuss how each security enhancement can be achieved.

\taggedpara{Revoking TA}
A service may continue to operate a malicious TA even after the Verifier has notified users of its adversarial behavior. 
    This is an issue, since some users may continue to use the malicious TA because they ignored, forgot, or simply did not receive the notification.

Although \sys itself cannot force the service operator to remove a TA, it can indirectly revoke the validity of TAs.
    Specifically, jointly working with the DR and CA, the Verifier can invalidate the TA domain and revoke its certificate, thus warning users not to connect with the TA.
    Note that, however, there may be some delay during this revocation process, mostly due to the DNS or CRL updates.

\taggedpara{Reducing CT Delay}
Even if \sys utilizes a custom CT monitor (as discussed in Appendix~\ref{sec:security-risk-delay}), we still face the delay of the CT log itself.
    We can reduce this delay by partnering with DRs.
    With the help of the DR, the Verifier can set the CAA DNS record~\cite{CAARECORD1, CAARECORD2} specifying which particular CA a TA could use.
    This allows the Verifier to force the TA to use a CA that registers its certificates on CT logs that exhibit short delay.
    Unfortunately, we have not yet found a CA that fits our criteria.

Instead of waiting for a CT log with a short delay to reach mainstream availability, we could also turn to CAs for their help to completely eliminate the delay.
    This is achieved by requiring the CA to verify the TA before issuing a certificate, rejecting requests if the CA detects that the TA is malicious.
    However, a clear drawback of this design is that no CA supports such functionality.

\subsubsection{Additional User-side Software}
\label{appendix:sec:additional-user-side-software}

The main motivation of \sys is to allow users to verify TAs without installing additional software on their platforms.
    However, some users may prioritize security and privacy over usability and functionality.

To meet these needs, the proposed system is designed to be capable of integrating with additional software installed by users. 
    For instance, users can directly verify remote attestation using software such as local proxies, custom SSL libraries, or browser extensions. 
    When contacted by the user, the TA provides $\evidence$ and $\pk$ in the response header, which can then be extracted using the above software and verified against the $\rv$ that was obtained in advance.

\subsubsection{Framing attacks}
\label{sec:framing_attacks}
A malicious Verifier can deceive a user by returning a falsified attestation verification result, leading the user to believe that a legitimate TA is malicious.
    Consequently, the user may choose not to interact with the TA, potentially due to this misinformation.
    Although this scenario is not part of our primary threat model, it remains a significant concern.
    From the TA's perspective, this attack is challenging to detect, as it is impossible to determine whether the user lost interest or was misled by a false verification result.
    The only way for the service operator to detect this attack is to pose as a normal user and visit the TA's website.
    Since the Verifier cannot distinguish between the user and the service operator, it will display the false verification result, enabling the service operator to become aware of the attack.

However, the malicious Verifier can still deny responsibility for the attack, claiming that the service operator is attempting to frame them.
    Unfortunately, the service operator must assume that Verifiers are legitimate, as they are naturally more powerful than the service (e.g., they can hold services accountable; see Appendix~\ref{sec:hold-responsibility}).
    To normalize this power dynamics, we would require the Verifier to operate within a TEE.
    This deters Verifiers from acting maliciously, as users and the service operator are able to obtain non-repudiable proof of the Verifier's malicious actions.

\subsubsection{Trusting Verifiers}
In this work, we assume that users do not trust verifiers.
    However, some users may opt-in to trusting them based on their reputation (e.g., operated by well-known companies or TEE vendors).
    This lowers the load on the user's side, as it alleviates them from checking the domain to prevent the attack described in Section~\ref{sec:security-analysis}.

\subsubsection{Detecting TAs running on vulnerable CPUs}
Typically, RA verification status reports include a list of information about the vulnerabilities patched in the CPU on which the report is running.
    Since CPU vulnerabilities are periodically discovered, it is crucial to promptly notify users whether the TA they are visiting is running on a CPU with the latest patches.
Although our threat model considers physical and side-channel attacks leading to these vulnerabilities as orthogonal to our system, we can extend the protocol to provide users with this information.
    Specifically, the Verifier should maintain a record of when the latest vulnerability for a TEE implementation was officially confirmed, as well as the date and time each TA registered.
    When contacted by a user during the communication phase, the Verifier should verify that the evidence was generated by a TA that is registered after the last known vulnerability was reported.
    If the verification is unsuccessful, this indicates that the TA has not been patched to mitigate the latest vulnerability, and the user should be alerted to this fact on the Verification Status page.

\subsection{Usability Analysis}
\subsubsection{UI fatigue}

For \sys to become widely adopted, it is important to minimize UI-related stress. 
    This section discusses the potential friction users may experience when using \sys. 
    The tasks required by users in \sys are: 
        (i) clicking the button shown on the TA website to open the Verification Status Page,
        (ii) checking TA information on Verifier sites, 
        (iii) acting on update notifications,
        (iv) acting on violation notifications, and
        (v) checking the domain of the pop-up page.
We emphasize that these tasks are unlikely to cause significant fatigue, as discussed below.

Tasks (i) and (ii) are necessary only when users access a TA for the first time. %
Meanwhile, task (iii) may occur more frequently. %
    Therefore, \sys provides an option for users to stop these notifications. 
    It is important to note that \sys remains secure even if users choose to do so, as we discuss in the next section.
Furthermore, we consider task (iv) to be rare in practice and, therefore, unlikely to cause significant stress.
Finally, although task (v) may have the largest burden, we can further reduce the load by optionally providing users tools that help them with domain checking.

While tasks (ii) and (iii) may require the user to manually check the TA source code which would be considered a high burden, especially for non-tech-savvy users, we discuss how we can minimize the load in the following section.

\subsubsection{Source Code Checking} 
\label{sec:code-auditing}

Reading source code can be a daunting task for non-technical users.
    This poses a challenge for \sys, which relies on users to review the source code of TAs.

To mitigate this issue, \sys adopts a strategy commonly used in Open Source Software (OSS) development.
    We assume that a sufficient number of tech-savvy and privacy-conscious users will review the TA source code, ensuring its integrity.
    This approach is not new; for instance, many users rely on the Linux kernel without examining its source code, trusting that developers will identify and address any issues.

To further simplify the process, users can subscribe to an OSS supply-chain security auditing service.
    This service can be integrated into the Verifier, allowing users to visually verify the legitimacy of the TA source code without needing to examine it in detail.

\subsection{ Openness Analysis }
\subsubsection{Browsers Requirements}
\label{appendix:browser-support}

To support \sys, browsers must provide certain essential functions.
    Specifically, they need to implement standard cryptographic communication protocols, including Web PKI, TLS, and CT.
    Additionally, browsers must also support several other key features, such as the Referer Header, Cookie, standard JavaScript runtime, and its APIs including \textit{window.open()}, Fetch API, Notification API, and Service Worker.

To verify the above assumption, we investigated browser functionalities using the MDN Web Docs. 
    The target browsers in our study were Chrome, Edge, Firefox, Opera, Safari, Chrome on Android, Firefox on Android, Opera on Android, Safari on iOS, and Samsung Internet. 
    \footnote{WebView on Android and iOS were not investigated as they are typically not considered as browsers.}
    Our investigation showed that every browser shown above supports all functions required by \sys, aside from CT~\cite{MDNCookieHT9:online,MDNSetCooki99:online,MDNRefererH75:online, MDNNotifica2:online, MDNServiceW50:online, MDNFetchAPI32:online, MDNWindowop32:online}.

To assess the extent of CT support, we conducted a manual investigation of CT adoption among the 10 browsers mentioned above.
    The results of this experiment are presented in Tables~\ref{table:ct-support-browsers} and \ref{table:ct-support-smartphone-browsers}.
    Our analysis reveals that a significant majority of browsers (8 out of 10) currently offer or plan to offer CT support.
    Although Firefox, Opera, and Samsung Internet currently lack CT support, Firefox has announced plans to integrate CT support~\cite{bugzilla_ct_policy:online} and Opera Developer (version 115.0.5314.0) supports CT~\cite{operadeveloper:online}.
    Furthermore, given the importance of CT in enhancing user security, we anticipate that the remaining browsers will likely adopt CT support in the near future.

\begin{table}[t]
    \centering
    \footnotesize
    \caption{Browser's support for Certificate Transparency}
    \begin{tabular}{lccccc}
         \toprule
       Browser  & Chrome & Edge & Firefox & Opera & Safari \\
     \midrule
       Version & 128.0.6613.137 & 129.0.2792.89 & 125.0.3 & 110.0.5130.23 & 18.0.1 \\
       Support CT? & Yes & Yes & No* & No\textsuperscript{$\dag$} & Yes \\
       \bottomrule
    \end{tabular}
    
    *: Plans to support CT in the future. \textsuperscript{$\dag$}: Supported on developer version.
    \label{table:ct-support-browsers}
\end{table}

\begin{table}[t]
    \centering
    \footnotesize
    \caption{Smartphone Browser's support for Certificate Transparency.}
    \begin{tabular}{lccccc}
         \toprule
       Browser & Chrome & Firefox & Opera & Safari & \makecell{Samsung\\Internet} \\
     \midrule
       Version & 129.0.6668.70 & 130.0.1 & 84.5.4452.81613 & 18.0 & 26.0.3.7 \\
       Support CT? & Yes & No* & No & Yes & No \\
       \bottomrule
    \end{tabular}
    
    *: Plans to support CT in the future.
    \label{table:ct-support-smartphone-browsers}
\end{table}

\subsubsection{ Additional Notification Methods }

Although \sys incorporates a notification mechanism, some users may miss notifications or prefer alternative schemes.
    Moreover, some browsers may not support certain notification mechanisms. %

To cater to diverse user needs, \sys can be extended to accommodate multiple notification mechanisms.
    For instance, we can opt for email notifications as an alternative.

\subsubsection{Accommodating Non-Mainstream Browsers}
\label{sec:non-mainstream-browsers}

Verifiers may opt not to support non-mainstream browsers, particularly those that have limitations such as the inability to provide user notifications or have installed certificates that the Verifier does not trust. 
    As a result, users who rely on these browsers will be unable to utilize \sys and verify the RA proofs from TAs that choose to use unsupported Verifiers. 
    This limitation raises concerns about openness.

To address this challenge, we introduce an option that enables anyone to act as a Verifier.
    This allows an interested party to develop and publish a Verifier specifically designed to support the previously mentioned browsers.
    We distinguish between two types of Verifiers: the \emph{Active Verifier} introduced in this scenario, and the traditional \emph{Passive Verifier} described earlier.
    As outlined in Section~\ref{sec:design}, a TA typically registers with a Verifier during the provisioning phase in \sys.
    However, the new approach reverses this process, enabling the Verifier to ``actively'' register the TA instead, allowing the Verifier to retrieve evidence from the TA, verify it, and store the relevant information in its databases.
    We envision Active Verifiers being operated by a multitude of entities; users using non-mainstream browsers and Nonprofit Organizations.

When operating active Verifiers, several key factors must be considered.
    Firstly, users must initially access the Verifier, which differs from the original system design where users first access the TA.
        This is because the TA is unaware of which Verifier is associated with it, preventing it from directing users to the correct Verifier.
        Note that an Active Verifier has a higher chance of being malicious than a Passive one since anyone can deploy a Verifier.
            We envision all the necessary countermeasures mentioned in Sections~\ref{sec:distributed-verifier} and \ref{appendix:ta-verifier} being deployed to mitigate this risk.
    Another factor is the option to hold TA responsible (see Appendix~\ref{sec:hold-responsibility}). 
        As mentioned earlier, holding TAs accountable for their actions requires authentication of their corporate status, which is not applicable in the case of Active Verifiers operated by potentially lesser-known third-party entities.
        Running Verifiers in TEEs can provide a solution to this issue (see Section~\ref{appendix:ta-verifier} for further discussion).
            By doing so, Active Verifiers can be trusted via the code they are running instead of their reputation.
            Specifically, during the corporation authentication process, TAs can also verify whether the Active Verifiers are running the expected code and, thereby ensuring their trustworthiness.

\subsubsection{Delivering Evidence via X.509 Extensions}

Some related works including RATS and SGX-RA-TLS explain that Evidence would be delivered via X.509 extensions.
    However, our implementation diverges from this design.
    This decision was motivated by our research, which revealed that some CAs, notably Let's Encrypt, remove non-standard data included in X.509 extensions.

We anticipate that this landscape will evolve soon, enabling the inclusion of Evidence in certificates.
    Notably, an IETF working group is currently discussing a standard for embedding Evidences within certificates~\cite{X509RA, ietf-rats-pkix-evidence-00}.
This approach offers a significant benefit: it enables open auditing of TAs. 
    By making all Evidence of a TA available on CT Logs, anyone can audit the TA. 
    This is particularly valuable when the Verifier is also a TA (as described in Appendix~\ref{appendix:ta-verifier}), as it allows anyone to audit the Verifier later.

\subsection{Load evaluation}

As one of the non-security goals defined in Section~\ref{sec:system} was minimal load on the Verifier, this section investigates how much load the Verifier experiences.
    Since measuring the load will depend heavily on the environment in which the Verifier is deployed, we conduct a theoretical analysis.

\subsubsection{Endpoints Access}
\label{sec:discussion:load:endpoint}

The Verifier API and Verification Result pages play a crucial role within the \sys ecosystem, as they provide users with essential information.
    Consequently, it is important to minimize the load on the server, given that user access tends to converge on this server.

We consider the load to be modest, due to two key factors.
    Firstly, user access to the Verifier is infrequent, occurring only when they initially visit a TA or when a TA notification expires.
    Secondly, the load generated by a single API access is negligible; it only takes 19.1 \si{\milli \second} for the total API access to finish (see Appendix~\ref{appendix:load} for details).

\subsubsection{CT Monitoring}

The Monitoring phase includes fetching and verifying certificates issued to the TA domain.
We argue that this process does not generate a significant load.
    This is because certificates can be fetched at a relatively slow rate (e.g., every 10 minutes), which is further mitigated by the inherent delay in the CT system.
Moreover, to minimize the load even further, we can utilize a CT monitor that offers a subscription-based notification system, such as the Meta CT Monitor, which alerts us of newly added certificates, thereby eliminating the need for frequent polling.

\subsubsection{User Notification}

Typically, push notifications are prone to high loads because they require establishing and maintaining connections with users behind Network Address Translation (NAT).
    However, as \sys leverages the W3C Push API~\cite{PushAPI57:online}, the Verifier can dispatch notifications without maintaining an open communication channel, thereby reducing the load.
Although the Verifier still needs to send individual notifications to each registered user, which could lead to a high load if the user base is large, the load is distributed across the entire \sys operation.
    This is because notifications are only sent when the Verifier detects a violation, which we assume occurs relatively infrequently, thereby minimizing the overall load.

\subsubsection{Code Building}

Code building undoubtedly constitutes one of the services that imposes a significant load on the Verifier. 
    Analogous to user notification, this process is not a frequent occurrence, thereby enabling us to distribute the load across the entire \sys operation.

However, note that a malicious TA operator may attempt to submit rapid-fire changes to the TA code in an effort to launch a DoS attack against the Verifier. 
    Although we do not consider this scenario within our threat model (see Section~\ref{sec:system}), the use of a standard rate-limiter, such as limiting the number of uploads via access tokens and other rate-limiting mechanisms~\cite{akama_scrappy_2024,nakatsuka2021cacti} will help mitigate this risk.

\subsubsection{Verifier-side storage}
\sys requires the Verifier to store information regarding the user for it to operate correctly. In this section, we go over two different types of information that the Verifier needs to store and show that it does not incur a substantial storage overhead.

\textbf{User subscription.}
The Verifier is required to maintain a subscription for each user, which is utilized during the notification process. This information is stored in a table named \texttt{subscriptions} (see Table~\ref{table:subscription-table} in Appendix~\ref{appendix:impl-details:dbs} for more details). If a large number of users are utilizing \sys, this subscription may impose a significant storage burden on the Verifier.

The size of a single record in the ``subscription'' table is a modest 313 bytes. 8 bytes (the maximum size consumed by an index in SQLite) is used for the index column in the record, rather than the actual size because the index size varies with the number of entries. Note that this measurement excludes SQLite's metadata and padding, which means that the actual size would be slightly larger in reality. Assuming that 100,000 users utilize 100 TA with \sys, we can calculate that the Verifier necessitates approximately 30.13 GB of storage capacity, which is not large for modern web services~\cite{StorageConsumption:online}.

\textbf{TA information.}
The Verifier is also required to store the TA information, which is stored in tables such as \texttt{ta\_codes} and \texttt{ta\_servers} (see Tables~\ref{table:ta_code-table} and \ref{table:ta_servers-table} in Appendix~\ref{appendix:impl-details:dbs}, respectively). Similar to the estimation for user subscriptions, we computed the amount of storage overhead this incurs. Our estimation showed that each TA information consumes a moderate amount of 6.265 KB. This consists of the measured 270 bytes of \texttt{public\_key}, 5,865 bytes of \texttt{quote}, and supposed 8 bytes of domain size, 33 bytes of Git Repository URL, 40 bytes Git Commit ID, and six 8 bytes of Integer IDs, two 1 bytes of Boolean Flags.

Assuming that 10,000 services are used and they updated their TA 1000 times, we can calculate that the Verifier requires approximately 62.65 GB of storage capacity. This is considered normal for modern web services~\cite{StorageConsumption:online}.

\subsection{Deployment considerations and incentives}

\subsubsection{Verifier operator incentives}
Verifiers play a crucial role in \sys and can be operated by various organizations with diverse incentives, such as:

\begin{itemize}
    \item TEE hardware vendors seeking to increase the appeal of their hardware;
    \item For-profit businesses that offer premium services, including higher levels of assurance, for a fee;
    \item Non-profit organizations that provide an alternative with a lower likelihood of conflict of interest.
\end{itemize}

The presence of multiple Verifiers is beneficial for the overall security of the system, as it enables service operators to swiftly switch to an alternative Verifier in the event of any issues.
Furthermore, this setup allows users to distribute their trust across multiple Verifiers, thereby preventing attacks, such as the one described in Appendix~\ref{sec:distributed-verifier}.

\subsubsection{Service operator incentives}
Supporting \sys also brings several benefits to services.
    Firstly, in terms of usability, \sys enhances the user experience by enabling users to utilize web services offered by TAs without the need to install additional software, thereby streamlining the process.
    Secondly, \sys prevents the installation of malicious software on the user side, providing users with peace of mind that their system's security is not compromised when visiting the TA's website.
    Additionally, services can promote their commitment to usability, security, and privacy by integrating \sys, which can serve as a valuable marketing advantage.

\section{Latency Measurement of Endpoints}
\label{appendix:load}

\textit{Evaluation settings: }
We measured the endpoint latency to estimate the load on the Verifier server, as discussed in Section~\ref{sec:discussion:load:endpoint}. 
    For this evaluation, we deployed the Verifier and accessed each endpoint.
    Table~\ref{table:latency-measurement} depicts the environment used for this evaluation. 
    Note that the Journal mode of SQLite is set to \texttt{wal} mode.
    Unless otherwise stated, all evaluations were run 5 times and numbers represent the average.

We conducted measurements under the following three settings.
The first setting is accessing the endpoints and recording the latency results from the console output provided by the Echo framework (denoted as Evaluation $\alpha$).
    In this setting, we accessed the Verification Result page and measured the latency of the 10 endpoints that are invoked immediately after loading the page.
    This represents the delay experienced by the user when utilizing \sys.
The second and third evaluations were conducted to measure the latency changes in relation to the number of records of logs stored in the database.
    The second setting (Evaluation $\beta$) changed the number of records for the target TA, while the third setting (Evaluation $\gamma$) changed the number of records for TAs other than the target TA.
    Note that for the second and third evaluations, we only measure the latency of the \texttt{/api/ta/:domain} and \texttt{/api/subscription} endpoints, since other endpoints do not get affected by the number of records in the database.

\begin{table}[t]
    \caption{The tools using the latency measurement}
    \label{table:latency-measurement}

    \centering
    \footnotesize
    \begin{tabular}{ll}
        \toprule
        Name &  Tools \\
        \midrule
        Cloud Provider & Microsoft Azure  \\
        Instance & DC1s v2 (1 vcpu, 4 GiB  memory)  \\
        Operating System  & Linux (Ubuntu 20.04)  \\
        Endpoints Access Tools (Evaluation $\alpha$) & Google Chrome \\
        Endpoints Access Tools (Evaluations $\beta$ and $\gamma$) & Chrome Developer Tools \\
        \bottomrule
    \end{tabular}
\end{table} 

\textit{Results:}
The results for Evaluations $\alpha$, $\beta$, and $\gamma$ are shown in Tables~\ref{table:load-endpoints}, \ref{table:load-shift-target-ta}, and \ref{table:load-shift-other-ta}, respectively. 
The results of Evaluation $\alpha$ show that most API endpoints display a delay that is in the order of several tens to hundreds of \si{\micro\second}, and the total latency is 19.1 \si{\milli\second}.
    We can see that the latency is dominated by the \texttt{/api/subscription} API.
        This is because this API requires the user to send a subscription token to the Verifier, which we hypothesize is influenced by the network delay.
    This can also be observed in Tables~\ref{table:load-shift-target-ta} and \ref{table:load-shift-other-ta}, as the \texttt{/api/subscription} API delay is roughly similar to the value shown in Table~\ref{table:load-endpoints}.
    Since these APIs are invoked only once for a certain TA per user and do not affect the critical execution path, we deem this delay acceptable for an average user.
Recall that Evaluation $\beta$ and $\gamma$ represent a less favorable condition where the number of records stored in the database is larger compared to that of $\alpha$.
    Even in these cases, we can observe that the increase in latency is negligible, mostly in the order of several tens of \si{\milli\second}.
Based on these results, we conclude that the latency of each endpoint is low, which, allows us to claim that the computational load is also minimal.

\begin{table}[t]
    \caption{The latency of the endpoints}
    \label{table:load-endpoints}

    \centering
    \footnotesize
    \begin{tabular}{clcc}
        \toprule
        Index & Endpoint & Static File & Latency [\si{\milli \second}] \\   
        \midrule
        1 & GET /app/verification-status/ & No & 0.0208 \\
        2 & GET /static/js/ui/app.js & Yes & 0.0258 \\
        3 & GET /static/js/notify.js & Yes & 0.0245 \\
        4 & GET /static/js/utils.js & Yes & 0.0190 \\
        5 & GET /static/js/react.js & Yes & 0.0111 \\
        6 & GET /static/js/ui/param.js & Yes & 0.0133 \\
        7 & GET /static/js/ui/table.js & Yes & 0.0268 \\
        8 & GET /api/config/subscription & No &  0.0237 \\
        9 & GET /static/js/sw/service-worker.js \footnote{we measured the latency by accessing this path in the browser.}& No & 0.0197  \\
        10 & POST /api/subscription & No & 18.4 \\
        11 & GET /api/ta/:domain & No & 0.517 \\
        \midrule
         Sum &  -  & - & 19.1 \\
        \bottomrule \\
    \end{tabular}
\end{table}

\begin{table}[t]
    \caption{Changes in latency when increasing number of records for target TA (Evaluation $\beta$).}
    \label{table:load-shift-target-ta}

    \centering
    \footnotesize
    \begin{tabular}{lll}
        \toprule
        Number of Records & \textit{/api/ta/:domain}[\si{\milli \second}] & \textit{/api/subscription}[\si{\milli \second}] \\   
        \midrule
        1 & 0.517 & 18.4  \\
        10 & 1.29 & 20.0 \\
        100 & 9.19 & 11.9 \\
        1000 & 72.3 & 11.4  \\
        \bottomrule \\
    \end{tabular}
\end{table} 

\begin{table}[t]
    \caption{Changes in latency when increasing the number of records for other TAs (Evaluation $\gamma$).}
    \label{table:load-shift-other-ta}

    \centering
    \footnotesize
    \begin{tabular}{lll}
        \toprule
        Number of Records & \textit{/api/ta/:domain}[\si{\milli \second}] & \textit{/api/subscription}[\si{\milli \second}] \\   
        \midrule
        1 & 0.433 & 12.6 \\
        10 & 0.461  & 11.7 \\
        100 & 0.499 & 25.9 \\
        1000 & 1.73  & 15.2 \\
        10000 & 10.7 & 23.7 \\
        \bottomrule \\
    \end{tabular}
\end{table} 

\normalsize
\raggedbottom
\clearpage

\end{document}